\documentclass[]{article}
\usepackage[utf8]{inputenc}

\usepackage[linktocpage=true]{hyperref}

\usepackage{xcolor}
\hypersetup{
	colorlinks,
	linkcolor={red!60!green},
	citecolor={red!50!blue},
	urlcolor={blue!80!black}
}
\usepackage{amsmath}
\usepackage[english]{babel}
\usepackage{graphicx}
\usepackage{makeidx}
\usepackage[usenames,dvipsnames]{pstricks}
\usepackage{setspace}
\usepackage{xcolor}
\usepackage{sectsty}
\usepackage{textcomp}
\usepackage{amssymb}
\usepackage{floatflt}
\usepackage{epsfig}
\usepackage{tocstyle}

\usetocstyle{standard}

\newcommand{\vn}{\vskip .5cm\noindent}
\newcommand{\nn}{\noindent}
\newcommand{\bs}{\begin{subequations}}
	\newcommand{\es}{\end{subequations}}

\makeatletter

\usepackage{amsfonts}
\usepackage{pdfpages}

\usepackage{setspace}

\usepackage{babel}
\usepackage{bookman,fancyhdr}
\usepackage{exscale}

\usepackage[linktocpage=true]{hyperref}

\makeatother

\usepackage{tocloft}

\title{\centerline{{\color{red!50!blue}\bf\Large Quantum Mechanical Reality: Entanglement and Decoherence}}}

\author{\centerline{{\color{red!50!blue}\bf\Large Avijit Lahiri\footnote{email: avijit.lahiri.al@gmail.com; blog (TacitKnowledge): tacit-views.blogspot.com}}}}

\date{}

\begin{document}

\maketitle

\newpage
\centerline{\bf Copyright}
\vskip 1cm
\centerline {\bf `Quantum Mechanical Reality: Entanglement and Decoherence' (an article)}
\vskip 2cm

\nn This work is licensed under the Creative Commons CC-BY-SA 4.0 License. To view a copy of the license, visit 

\vn https://creativecommons.org/licenses/by-sa/4.0/

\vskip 3cm

\nn Kolkata (India): 20 July, 2023.

\vskip 1cm

\nn avijit.lahiri.al@gmail.com

\newpage

\begin{abstract}
	
\nn We look into the ontology of quantum theory as distinct from that of the classical theory in the sciences. Theories carry with them their own ontology while the metaphysics may remain the same in the background. We follow a broadly Kantian tradition, distinguishing between the noumenal and phenomenal realities where the former is independent of our perception while the latter is assembled from the former by means of fragmentary bits of interpretation. Theories do not tell us how the noumenal world is constituted but are conceptual constructs applying to {\it models} generated in the phenomenal world within limited contexts.

\vn The ontology of quantum theory principally rests on the view that entities in the world are pervasively correlated with one another not by means of probabilities as in the case of the classical theory, but by means of probability {\it amplitudes} involving finely tuned {\it phases} characterizing the oscillatory behavior of quantum mechanical states. The amplitude-dependent correlations (quantum {\it entanglement}) exist over and above the classical ones expressed in terms of probabilities. While the classical correlations are essentially local in nature, quantum correlations are {\it shared globally} in the process of environment-induced {\it decoherence}. The decoherence is an effectively random process that removes local correlations in the course of global sharing of entanglement---the removal being especially manifest in the case of systems that appear as classical ones. It is this aspect of the decoherence process that makes the so-called measurement postulate (one relating to {\it wave function collapse}) cohere with the rest of the principles of quantum theory where the latter implies a Schr\"odinger type time evolution of quantum mechanical systems. 

\vn The mathematical basis of the quantum correlations consists of the description of pure states of a system in terms of vectors in a linear vector space (a {\it mixed} state appears as an admixture of pure states with some probability distribution associated with it) and, in addition, the description of (pure) states of composite systems as vectors in the {\it product space} arising from the component sub-systems. 

\vn The crucial aspect of the decoherence process, of significance in the context of the apparent incompatibility of the measurement postulate with the unitary Schr\"odinger evolution, relates to the fact that it is almost {\it instantaneous} in the case of a classical object (one that is nevertheless amenable to a quantum description). Indeed, the decoherence time is, in all likelihood, of the order of the {\it Planck scale}, being driven by {\it field fluctuations} in the Planck regime. This points to factors of an unknown nature determining the finest details of the decoherence process since Planck scale physics remains an obscure terrain.

\vn In other words, quantum theory is, to all intents and purposes, in the need of a radical revision, in keeping with the fact that {\it all} theories are defeasible and need revision as our domains of experience expand and get realigned in a complex manner. The context within which quantum theory (and quantum field theory too) is defined is set precisely by the Planck scale, across which a novel theoretical framework is likely to emerge. However, as in the case of theory revisions in general, that emerging theory will stand in an asymmetric relation of {\it incommensurability} with the present-day quantum theory, where the concepts of the latter will be comprehensible in terms of those of the former, but the converse will not hold.  

\end{abstract}

\newpage

\section{Introduction and outline}\label{preamble-sec}

\nn We begin this article by making a few general observations on reality and our conception of it. This follows a broadly Kantian tradition where the {\it noumenal} and the {\it phenomenal} worlds are clearly distinguished, being two aspects of the same reality. We follow a naturalized version of this tradition where the infinite {\it complexity} of the noumenal reality is seen as a barrier against the possibility of grasping or comprehending it {\it as a whole}, in virtue of which we need to assemble fragmentary and patchy interpretations of it, thereby generating the phenomenal world---our conception of reality. We briefly outline (section~\ref{mind-sec}) how the human mind confronts and connects with reality, building {\it theories} of how the world behaves. Theories are applicable to {\it models} pertaining to the phenomenal world and are conceptual constructs, without these being approximations to some all-embracing and foundational `law of nature'. With accumulating experience in ever broader contexts, our theories of the world get revised,---at times radically---with emerging theories applying to ever more novel contexts.

\vn In the next section (sec.~\ref{problems-sec}) we look at the purported `problems' of quantum theory as compared with our familiar classical view, and also indicate the sense in which quantum theory can be seen to be in need of a revision. 

\vn We view quantum theory and its possible future revision in the general context of how we build and revise theories and, as mentioned, look at its ontology within a broadly Kantian metaphysics. Quantum ontology differs fundamentally from its classical counterpart in the description of the state of a system as a vector in a linear vector space where the superposition principle applies, and where the idea of `pre-existing' values of physical properties in any given state is denied. This is commonly considered to indicate a fundamental lack of `objectivity' (section~\ref{objectivity-sec}, section~\ref{quantum-revision-sec}) in quantum theory, a perception responsible for massive efforts aimed at removing this purported deficiency. However, every emerging theory of the phenomenal world carries its own ontology when compared to previously existing ones since, in contrast to metaphysics, our ontology is inextricably linked with our epistemic framework, and quantum theory is no exception. The ontology specific to it (see sec.~\ref{ontology-sec} later in this essay) can be described as one involving a pervasive phase-dependent correlation (or, more specifically an amplitude-dependent one instead of a classical probability-based correlation) among entities of the world---namely, the quantum {\it entanglement} as distinct from classical correlations between pre-existing values in terms of probabilities. What is more, entanglement between quantum mechanical systems is inextricably linked with {\it decoherence} (sec.~\ref{decoherence-sec}), involving a {\it global} entanglement sharing between environmental degrees of freedom. All this distinguishes the ontology of quantum theory from the classical one.

\vn\begin{quote}\begin{scriptsize} Two other respects in which quantum ontology differs essentially from the classical one relate to {\it fields} as the basic entities of the world and to the relevance of {\it internal degrees of freedom}, such as the spin, in describing the states of particles, the latter being, in turn, states of underlying fields. The idea of the internal degrees of freedom and that of fields are related ones. However, this will not be of central relevance in the present discussion.\end{scriptsize}\end{quote}  

\vn Quantum theory, however, lacks in internal {\it coherence} (section~\ref{objectivity-sec}) since the {\it measurement postulate} is incompatible with evolution principle implied by the Schr\"odinger equation. It is this same measurement postulate, along with the superposition principle and the feature of entanglement, that leads to paradoxical features such as non-locality and contextuality of hidden variable theories aiming at an `objective' interpretation of quantum measurement results. 

\vn The next section (sec.~\ref{decoherence-sec}) and its subsections deal with the role of {\it environmental decoherence} in quantum measurements. The theory of decoherence is seen as one that rids quantum theory of the measurement postulate by providing an explanation (of sorts) of the same in terms consistent with the Schr\"odinger equation, where one makes use of the inordinately short time scale of environmental decoherence of classical systems---assuming the decoherence time to be zero in a limiting sense leads to the measurement postulate. However, the detailed theory of environmental decoherence is likely to lead to the consideration of processes across the {\it Planck scale}, which is where quantum theory (including the quantum field theory) is likely to get revised in a radical manner, leading to a novel theory, carrying with it a novel ontology too.

\vn These considerations will be followed by section~\ref{reality-sec} where we revisit the issue of ontology of quantum theory and sum up with a number of observations on the nature of reality that quantum theory portrays, indicating the {\it context} in which that portrayal is an effective one. Like other theories in science, quantum theory is likely to be revised as that context is changed and novel phenomena are unearthed beyond the so-called Planck scale---a terrain that a detailed consideration of the decoherence phenomenon is expected to lead us to. 

\vn This essay is for a broad audience and avoids technical considerations in quantum theory and the theory of environmental decoherence. The necessary technical notions that constitute the background of section~\ref{decoherence-sec} have been listed (without explanation) in sec.~\ref{technical-sec}, along with reference to literature that will serve to elucidate these notions. 

\section{Reality and the human mind}\label{mind-sec}

\vn\begin{quote}\begin{scriptsize}As background to the present section, see~\cite{Lahiri1}, and references therein. For parts of the present section (sec.~\ref{mind-sec}) and also of sec.~\ref{reality-sec}, refer to~\cite{Baggott1}, \cite{Baggott2}.\end{scriptsize}\end{quote}

\nn\begin{enumerate}

\item We begin by briefly stating our position on metaphysics and ontology. The former is essentially the one espoused by Kant, with the added ingredient---an essential component of the world-view adopted in this paper---based on {\it complexity of reality and of entities that reality is made up of}. It is the same complexity that also has its indelible effect on our epistemology---how we comprehend reality. These issues will be addressed by way of making only bare-bones statements that will tell us where we stand in matters of foundations.

\item Considerations of complexity serve to provide a concrete foundation to Kant’s metaphysics and ontology, and are meant to naturalize the world-view he propounded. The naturalist trend in philosophy calls for a revision of the philosophical world-view in keeping with the progression of scientific theories through stages. Kant, in a sense, was a naturalist in his time and his views have a general affinity to an interpretation in terms of insights gained from complexity theory over the last fifty years.

\vn\begin{quote}\begin{scriptsize} The science of complexity is of relatively recent vintage, that includes the theory of {\it algorithmic} complexity but is itself of a much broader scope. For an introductory and general overview, see~\cite{Simon}, \cite{Holland}, \cite{Thurner}, \cite{Fuchs}, \cite{Lahiri1}.\end{scriptsize}\end{quote}

\item We distinguish between the reality existing independently of our conception of it, and the one we assemble by means of our interpretations of the myriads of signals received incessantly from that mind-independent reality. We refer to these two aspects of reality as, respectively, the {\it noumenal} and the {\it phenomenal}, following the Kantian tradition.  The fundamental idea underlying our metaphysics is that, in keeping with the complexity of entities in the phenomenal world, the noumenal world is a pervasively complex one, where the distinction between the reality as a whole and the entities belonging to it is ruled out in principle, since that distinction holds only within limited ranges of space and time. The most basic characteristic feature of a complex system is that all its numerous constituent entities are correlated with one another in an intricate manner, in virtue of which the said constituents can be distinguished from one another (are {\it decomposable}) only conditionally and contextually.

\item The human mind takes cognizance of reality by an {\it active} process whereby signals received from the world are integrated with {\it stored} interpretations resulting from past experience, thereby generating interpretations of current perceptions, and anticipations of future phenomena. Since the noumenal world exists as a whole and in itself, with intricate correlations among all the entities making it up, our phenomenal perceptions are by their very nature partial and contextual, in virtue of which our interpretations of reality are fragmentary and patchy too. At any given time, all these fragmentary pieces of interpretation are assembled to form the phenomenal world which constitutes {\it all} that the noumenal world offers us. The phenomenal, in other words, is a {\it projection} from the noumenal, which is infinite-dimensional  (having an infinite number of independent qualities required for our notional conception of it) and is a complex whole---it is the source of all our perceptions, but cannot be accessed or interpreted in its entirety, which rules out the backward projection from the phenomenal to the noumenal except by way of a meta-induction.

\vn\begin{quote}\begin{scriptsize}Interestingly, whie the noumenal world is conceptually inaccessible (all our concepts pertain to the phenomenal), it is certainly affected by our {\it actions}. When we act upon the world ({\it punching a table-top in frustrated anger}, {\it applying a magnetic field to a moving electron}), it is the noumenal world that is impacted, while the results of that action are once again apparent as the phenomenal. The latter is nothing but the apparent face of the former---the face that we access and comprehend conceptually.\end{scriptsize}\end{quote} 

\item The phenomenal world appears to be decomposable in terms of entities that have an existence of their own, though the existence of an entity (a `being') is manifested only by means of its interactions with other entities in it resulting in its evolution (`becoming') over time. The two aspects of being and becoming are separated from each other but that too only conditionally and contextually. We conceive of an object only within some given context set by our mode and means of observation and our current conceptual state of interpretations accumulated over the past. Additionally, the world we perceive is characterized by a vast hierarchy of {\it scales} in time and space, and the entities and their evolution captured in our mind are all conditioned by the scales in which these are perceived. It is precisely this conditional and contextual aspect of our perception and interpretation that lies at the basis of the way the phenomenal world is assembled: the same noumenal reality that exists in itself as an infinite and integral whole, when perceived conditionally and  contextually, gives rise to our impression of entities (objects and experiences) existing apart from one another and interacting between themselves, causing their evolution in time. In this sense, the noumenal and the phenomenal are not two distinct worlds, but one and the same world---the phenomenal being the perceptual and conceptual reconstruction of the noumenal, fragmentary and partial in nature. It is the nature of the reconstruction that is the most interesting and central question of epistemology. Our phenomenal world gets assembled as we attach {\it meaning} to the myriads of signals received from the noumenal reality, the signals themselves being devoid of meaning.

\item The human mind confronts and connects to the world by means of two modes of `logic'---the logic of {\it affect} and the logic of {\it reason} (the {\it implicit} and the {\it explicit}) where, once again, the two can be distinguished only notionally---both originate from underlying processes in the brain and have subtle inter-dependencies. Affect is generated mostly below the level of awareness by the activity of the affect network in the brain and is responsible for the vast and pervasive set of {\it preferences} lodged in our mind, ever more of those being produced incessantly in the course of accumulating experience. Reason, on the other hand, operates at the conscious level on the basis of {\it relations} among entities of the world perceived without overt reference to our self-based preferences. It is on the basis of the strange, intimate, and paradoxical partnership between the two that we engage in processes of making {\it decisions} and {\it inferences}---these being of essential relevance making possible our eventful journey through life.

\vn Interpretation--decision--inference, all these are made possible as our perception interacts with a vast store of preferences, beliefs, dispositions, and intents in our mind, all linked with affect and reason.

\item All these, finally, lead to {\it theories} of nature. Theories are made up of bundles of beliefs, some well justified and some not so much so. Theories enable us to explain the world---the phenomenal world, that is,---and to predict how it behaves. Theories apply to models, i.e., parts of reality scooped out from the rest where the effect of the latter is taken into consideration (in an approximate manner) by specifying the context in which a model is set up. A model, in a sense, is an idealization, that is no substitute for reality but one that nevertheless bears a very definite relevance in our phenomenal experience. The efficacy of a theory arises within certain spatial and temporal scales set by the context in which it is set up. A theory is based on certain regularities found in our phenomenal experience and aims at explaining those regularities in terms of a relatively few chosen concepts along with consequences following from those concepts.

\item Theories are often said to be expressions of `laws inherent in nature', and successive revisions of theories are interpreted as closer and closer approaches to some ultimate and fundamental Law of Nature. According to this view, the theories that emerge in our conceptual world correspond more and more closely to laws of nature that have an ‘objective’ existence out there in reality, waiting to be discovered by us. Stated in terms of the framework we outline above, this represents a backward projection from the phenomenal world to the noumenal while only forward projections from the latter to the former make sense. Theories are meaningful only within the confines of the phenomenal world and are conceptual constructs for explaining the regularities in it and are not ‘objective’ in the sense of corresponding to principles inherent in the workings of the mind-independent noumenal reality.

\vn\begin{quote}\begin{scriptsize}We do not know what the noumenal reality is `really' constituted of and how it evolves---all our knowledge is confined only to the phenomenal, even as the noumenal is the ultimate source of that knowledge. But we can guess, learning from our experience of complex systems and the vast literature now accumulated on such systems. Complex systems evolve through stages of {\it self-organized complexity}, passing through alternating regimes of stability and instability across a large hierarchy of spatial and temporal scales. During such evolution, a complex system gets caught on islands of regularity within a background of unstable and irregular behavior. All our inferences and theories pertain to such islands of regularity, where the inferring mind succeeds in capturing numerous apects of that regularity.\end{scriptsize}\end{quote} 

\item Successive revisions of theory do not correspond to a more and more close approach to rules residing objectively in Nature, but represent the emergence of ever new concepts and beliefs as we attempt to explain our accumulating experience in sporadically changing contexts. The revisions do not correspond to better and better approximations to truth but to distinct {\it perspectives} in which nature is viewed---distinct perspectives, one may say, that we adopt while comprehending the noumenal reality. Our world-view is generated as a patched-up assembly of all these distinct perspectives. Theories emerging in successive revisions carry a definite sense of progression when compared among themselves but are, by the very nature of things, fundamentally incapable of telling us how the noumenal reality is constituted---this is analogous to the impossibility of guessing a geometrical structure in an infinite-dimensional space from a finite dimensional projection of it (or even guessing at a three-dimensional structure from a number of two-dimensional or one-dimensional projections). What is more, theories, in a very definite sense, can be said to be {\it incommensurate} with respect to one another. We will come back to the issue of theories and their relevance in the particular context of quantum theory later in this essay (sec.~\ref{reality-sec}).

\end{enumerate}
 
\section{Quantum theory and its purported problems}\label{problems-sec}

\subsection{Introduction: objectivity and internal coherence}\label{objectivity-sec}

\nn Quantum theory, despite its enormous success, is thought to be ‘problematic’ by numerous commentators, including scientists and philosophers. This view is almost universally a consequence of the fact that the theory is seemingly at odds with the requirement of ‘objectivity’. A theory is said to satisfy this requirement if its basic principles and predictions refer to {\it true} properties of entities of the world regardless of the process of observation or measurement that an entity may be subjected to. Quantum theory is perceived to be at variance with this philosophical requirement even though one recognizes that the philosophy in question has originated from grounds of pre-quantum, i.e., the so-called classical, theories. The sense of a deficiency in quantum theory is heightened by the fact that a commonly accepted point of view of `understanding’  both classical and quantum theories within a single framework has not emerged. By and large, the perceived deficiency of quantum theory is associated with a lack of comprehension as to how the theory relates to the real world that exists independently of the human mind. In this context, processes of observation and measurement are deemed to be, at least partially, of a subjective nature. 

\vn But theories themselves pertain to the phenomenal and not to the noumenal world. Their purpose is not to provide an accurate description of the latter but to explain and predict within specified contexts in the former. In this respect, quantum theory does not differ from other theories with explanatory and predictive value. The classical theories differ from the quantum theory in the sense that their explanatory and predictive power is perceived to result from the ability to provide a true description of the mind-independent universe. It is quantum theory that for the first time clearly exposes the divide between the two---the power of a theory to explain and predict within the phenomenal world and the ability to provide a true description of the mind-independent noumenal world. The latter is an {\it infinitely connected} and {\it complex} whole, and a conditional and contextual separation of a part from the whole immediately takes us away from that whole into the phenomenal universe captured and assembled in our conception. 

\vn The ‘no-miracle’ argument is often produced to justify the claim that explanatory and predictive power has to result from the ability to provide a true description of how the world operates. But this confounds the distinction between the noumenal and the phenomenal. Theories are certainly meant to provide correct descriptions of the world, but that world is the phenomenal one where all concepts and all consequences of theories are meaningful only conditionally and contextually. It does not make sense to speak of truth regardless of context because there is no way to access such Platonic truth. In this sense, quantum theory is a theory about our phenomenal world, the world made up of all our experiences, conceptions, beliefs, and abstractions. All the principles defining quantum theory and all consequences following from these hold only within a context, namely, within the non-relativistic ‘microscopic world’, as it is commonly referred to. Quantum field theory---the extension of non-relativistic quantum theory into the realm of relativistic space-time---holds within the context set by the so-called Planck scale which is also the ultimate horizon limiting the validity of non-relativistic quantum theory. The question now arises as to whether quantum theory is `complete’. The answer to this question is clearly in the negative even before we enter into a concrete  comparison of quantum theory with observations, because quantum theory, like any other theory, applies only to the context mentioned above.  Completeness can only be conceived of with reference to the noumenal world which, in any case, is inaccessible as it is---what is accessible is a collection of fragmentary and partial perspective views of it captured in our conceptual world.

\vn On the other hand, one can very well ask as to whether quantum theory is `consistent' or `coherent'. No inconsistency has ever been detected in quantum theory during a period spanning over a century. However, the theory is not a coherent one---the measurement postulate does not cohere with the rest of the principles defining quantum mechanics. Indeed, it is at odds with the evolution principle implied by the Schr\"odinger equation though, to all intents and purposes, it correctly explains all experimental observations within the domain mentioned above.

\vn While no theory, including quantum mechanics, is unconditionally objective in the sense of providing a true description of the world and asserting that an entity belonging to the world is characterized by a pre-existing value of every observable quantity in any given state of it, one has to explain why classical physics appears to satisfy the requirement of objectivity. This distinction between the quantum and the classical theories relates to the fact that quantum theory carries with it an ontology fundamentally distinct from that of classical mechanics (see sec.~\ref{quantum-revision-sec} below as well as sec.~\ref{ontology-sec}).

\subsection{In what sense does quantum theory need a revision?}\label{quantum-revision-sec}

\nn The fundamental divergence between the classical and quantum mechanical descriptions of systems stems from the way the state of a system is represented in these two. In the classical theory, the state (a pure state, that is; mixed states correspond to probability distributions over pure states) is represented by a point in the appropriate phase space, where each of the observable quantities of the system under consideration is characterized by some specific value. If an observation or measurement pertaining to an observable quantity is made on the system in some specified state, then the value corresponding to that state (referred to as a ‘pre-existing’ value) will be obtained as the result. This is far from the case in the quantum theoretic description of the state of a system where, generally speaking, the result of measurement of an observable in a pure state (a mixed state is once again given by a probability distribution over pure states) does not correspond to a pre-existing value but obeys a probability distribution over a set of values characteristic of the observable (the eigenvalues of the latter)---it is the probability distribution that is characteristic of the state. This, indeed, is the foundational ontology of quantum theory that gets expressed by its mathematical formalism where states are represented by vectors in a linear vector space (technically, a Hilbert space; also termed the `state space') and observable quantities by linear Hermitian operators. The expectation value of an observable in any given state of a system is given by the Born rule, which is related to the (infamous) measurement postulate that states that, upon an observable having been measured in any given state and one of its eigenvalues having been obtained as the measured value (with a probability given by the Born rule), the system `collapses' into the corresponding eigenstate (assuming the eigenvalue to be non-degenerate for the sake of simplicity). 

\vn\begin{quote}\begin{scriptsize}See sec.~\ref{ontology-sec} for further considerations on the ontology of quantum theory.\end{scriptsize}\end{quote}

\vn Such collapse, indeed, is inconsistent with the Schr\"odinger equation, the foundational equation describing the evolution of a state in terms of its Hamiltonian operator. The measurement postulate also leads to problems with objectivity since the value of an observable that is obtained in a measurement cannot be interpreted as a pre-existing one. The features of non-locality and, more generally, contextuality, of classically objective theories designed to account for quantum mechanical observation are ultimately related to this perceived deficiency of quantum theory. And it is in this respect that quantum theory stands in need of a revision. However, as we see later in this essay, such a revision is going to be no easy thing to come by, because it is likely to require the crossing of a fundamental barrier in our domain of experience whereby our entire conceptual world will be poised to undergo a remarkable transformation.

\section{Environmental decoherence and the measurement problem: a brief overview} \label{decoherence-sec}

\subsection{Necessary technical notions}\label{technical-sec}

\nn It is in this context that the theory of environmental decoherence assumes central relevance, since it provides a justification (of sorts) of the measurement postulate, exposing how the measurement process conforms to the Schr\"odinger evolution principle in a limiting sense. The necessary basic notions are those relating to the following: (a) the superposition principle and the possibility of expansion of any given state vector as a sum over eigenstates of any chosen observable; (b) the notion of pure and mixed states of a system and of the density matrix that provides a common description of the two types of states; (c) the representation of the state of a composite system (say, C) made up of sub-systems (say A, B---this is the case of a {\it bipartite} system; more generally, multi-partite systems can be considered) by a vector in the {\it product space} of the state spaces of A, B; (d) the {\it relative entropy} ${\cal E}(\hat\phi||\hat\psi)$ between two given states  ($\hat\phi, \hat\psi$; we adopt a commonly accepted notation and consider mixed states here for the sake of generality) of a system; this is a measure of the {\it separation} between the two states (though not in the strict sense of a metric; other measures also exist) under consideration; (e) a {\it separable} state of a composite system (say, C referred to above), i.e., one that can be expressed as a sum over states, each term in the sum being, generally speaking, a product of mixed states of the sub-systems; (f) an {\it entangled} state---one that cannot be expressed in the above form of a separable state; (g) the {\it reduced} states (referred to subsystems A, B) of any given state of a composite system (C in the present instance; the reduced state of A, say, is the one that gives correct values in the measurements of all {\it local} observables, regardless of results of measurements on B); and (h) the entanglement measure of a state of a composite system---there exists alternative measures, equivalent in spirit---defined as the separation from its {\it nearest separable state}.

\vn\begin{quote}\begin{scriptsize}The above notions are defined and explained in~\cite{Nielsen}, \cite{Vedral}, \cite{Vedral-Plenio} among many other excellent texts on quantum theory. The present essay is meant for a broad audience and avoids the use of technical notations to the extent possible; even a superficial familiarity with the above notions will be enough to make this essay intelligible to the discerning reader. Excellent background to the theory of environmental decoherence, along with reference to relevant literature is to be found in~\cite{Schlosshauer}. The following account of the role of environmental decoherence in explaining a number of issues in quantum measurements and more generally in quantum theory itself, is based on~\cite{Lahiri2}, \cite{Lahiri3}, while ~\cite{Vedral} is very similar in spirit. \end{scriptsize}\end{quote} 

\vn A separable state pertaining to sub-systems A, B involves only a {\it classical} correlation between states of A, B. The measure of the classical correlation in such a state is given by the separation from the direct product of the respective reduced states of the sub-systems. The entanglement measure or (the {\it quantum} correlations) of a state of C represents a non-classical correlation that exists over and above the classical correlations in the nearest separable state.

\vn Entanglement between quantum mechanical systems arises in virtue of interactions among those in the course of their Schr\"odinger evolution and gets {\it shared} with systems with which these interact at subsequent times. In a manner of speaking, entanglement between systems involves a delicate {\it phase correlation} between these---in this sense, one distinguishes between {\it amplitude-based} quantum correlations and {\it probability-based} classical correlations, where the latter can be expressed in terms of pre-existing values of observables in the case of classical systems (refer to sec.~\ref{classical-sec}). Phase correlations are fragile in nature---those between systems A and B can get reduced as either of these two interacts with a third system.

\subsection{The measurement problem}\label{measureprob-sec}

\nn The so-called `measurement problem' is essentially the one of explaining away the incompatibility of the measurement postulate with the rest of the principles of quantum mechanics and of justifying the context-dependence of the results of quantum measurements. These issues were extensively addressed by Niels Bohr, especially in his exchanges with Einstein, and were later sought to be formulated in technical terms by Von Neumann.

\vn Bohr underlined the importance of the `measurement interaction' between the quantum mechanical system on which a measurement is made, and the measuring apparatus used, and also of the essentally {\it classical} nature of the latter. Von Neumann explicitly referred to the entanglement between the measured system and the apparatus established by means of the measurement interaction, and formulated his `measurement model' that served as the basis of subsequent developments(~\cite{Schlosshauer}). However, the classical nature of the measuring apparatus was not explicitly made use of in Von Neumann's model, though Neumann made major contributions to the quantum mechanics of classical systems and produced (in parallel with Koopman) the formulation of classical mechanics in formal analogy to quantum principles.

\subsection{Environmental decoherence of states of classical objects}\label{classical-sec}

\vn Finally, the measurement postulate was sought to be explained on the basis of {\it Decoherence theory}, pioneered by Zeh, Zurek, Paz, and other contemporary exponents (see~\cite{Schlosshauer}). Briefly, a quantum mechanical system interacts incessantly with entities belonging to its environment (commonly, the atmosphere or residual molecules in an evacuated chamber, but {\it additionally}, on ubiquitous {\it field fluctuations}), as a result of which its state gets entangled with those of the environment. A major consequence is that the quantum correlations between systems (produced in virtue of interactions between them) undergoes a process of decay (the decoherence process) as one or more of these are subjected to interactions with the environment, since the latter is essentially in the nature of a {\it random} process, involving a {\it global entanglement sharing}. What is of major relevance in this process of environmental decoherence is its {\it time scale}. In particular, the environmental decoherence of {\it classical} objects can be, to all intents and purposes, {\it instantaneous}. 

\vn More explicitly, a classical object is characterized by a relatively few {\it macroscopic} variables (the shape and size of a solid body, for instance) of a collective nature, along with an enormous number of {\it microscopic variables}, and a pure state of the system at any instant---a vector in a state space of an equally enormous number of dimensions---can be specified in terms of the values of all these variables. Correspondingly, the state space can be looked upon as a direct product of a space corresponding to the collective variables and one corresponding to the microscopic variables. Imagining a superposition of such states with two given sets of values of the macroscopic variables (there may correspond a large number of values of the microscopic variables in the superposition), the relative phases of the superposed states, which give rise to {\it interference terms} are responsible for quantum correlations among these states. The state in question undergoes decoherence under the influence of the environment and quickly gives rise to a mixed state, the latter being a mixture with all possible values of the microscopic variables and with the two sets of macroscopic variables mentioned above (these remain intact owing to approximate conservation principles), where the phase correlations are completely erased now under the impact of the enormously large number of environmental degrees of freedom. One thereby ends up with a mixed state of the system involving two well defined sets of values of the collective variables and all possible values of the microscopic ones. This is the sense in which a superposed state of a classical object can be said to go over to a mixed state where all correlations are of a classical nature since quantum correlations are completely erased by the environmental effect.

\vn\begin{quote}\begin{scriptsize}The macroscopic variables are robust because of their collective nature and because of approximate conservation principles. \end{scriptsize}\end{quote} 

\vn The {\it time scale} of the above decoherence process involving a classical object is of great relevance. Since it requires only a loss of phase correlations and since both the sets of microscopic variables characterizing the classical object and the environmental variables are enormously large in number, the decoherence process is an exponentially fast one (\cite{Schlosshauer}). Concrete estimates of the decoherence time for realistic systems are few in number and it appears that the environmental variables relevant for the decoherence process correspond to {\it field fluctuations}, with a role more significant than environmental particles scattered from the object in question. Order of magnitude estimates show that the time scale of the decoherence process of a classical system can be even smaller than the {\it Planck time scale} (in this context, refer to~\cite{Petruzziello}, \cite{Arzano}, \cite{Diosi}). In other words, a rigorous analysis of the process is likely to open the door to a space-time domain that has so long remained unexplored. As mentioned above, it is precisely this feature of the process of environmental decoherence of classical objects that is likely to provide an explanation of the phenomenon of wave function collapse. In this context, it is important to note that the decoherence process, while being effectively random in nature, is one that follows the Schr\"odinger evolution principle, where the evolution can be described in terms of a reversible untary operator acting on the state space. The seemingly irreversible nature of the process of decay of phase correlations owes its origin to the global sharing of correlations among the environmental degrees of freedom (oscillatory modes of fields ubiquitous in nature).

\subsection{The measurement interaction and the measurement process}\label{measurement-interact-sec}

\vn Generally speaking, a quantum measurement is aimed at making a quantum mechanical system interact with an appropriate measuring apparatus so as to obtain the `measured value' of some observable pertaining to the system, where the measured value does not refer to any pre-existing value of the observable. In order that the measurement be possible, the apparatus has to be an appropriate one, depending on the observable being measured. For instance, the apparatus measuring the spin of a particle along any given direction in space cannot be the same as the one measuring the spatial location of the particle on any specified axis. Depending on the observable quantity measured, there exist apparatus states with which the eigenstates (belonging to the observable in question) of the measured system get correlated by the interaction between the system and the apparatus.

\vn\begin{quote}\begin{scriptsize}The present section is a condensed version of a brief note (\cite{Lahiri2}) based on more explicit technical considerations in~\cite{Lahiri3}. \end{scriptsize}\end{quote}

\vn As emphasized by Bohr, the apparatus used in a quantum measurement is generally of a classical nature, and it is this specific feature of the measurement process that is responsible for all its paradoxical features. In accordance with what has been outlined above, the process of environmental decoherence operates incessantly on the apparatus, and any pure state of the latter is instantaneously converted into a mixed state, where a mixed state is associated with some specific value of a macroscopic apparatus variable referred to as a {\it pointer variable} that depends on the constitution of the apparatus. Since the apparatus is chosen in accordance with the observable (of the measured system) of interest, it follows that the environment always acts on the apparatus so as to cause the latter to exist in one of a set of mixed pointer states, each such state being characterized by some specific value of the relevant pointer variable. Depending on what observable is being measured, the apparatus has to be an appropriate one in the sense that its set of pointer states has to correspond to the eigenstates of that observable.

\vn According to the Von Neumann scheme of quantum measurements, the interaction between the measured system and the apparatus (the `measurement interaction' as it is referred to) results in a system-apparatus {\it entangled} state where eigenstates of the measured observable get associated with the corresponding pointer states of the apparatus. It may be noted that the evolution of the state of the composite system made up of the measured system and the apparatus leading to the entangled state mentioned above occurs in parallel with the process of environment-induced decoherence of apparatus states so that, at every instant, the apparatus states involved in the measurement interaction are precisely the pointer states mentioned above.

\vn In the Von Neumann scheme, both the measured system and the measuring apparatus are treated quantum mechanically, but the latter has a special status in virtue of its classical nature since, at each and every instant, all quantum correlations between the apparatus subsystems get erased in virtue of the process of environmental decoherence. In other words, the apparatus is a limiting instance of a quantum mechanical system. However, though the apparatus subsystems have no quantum correlations between them, the measurement interaction does introduce quantum correlations between the relevant eigenstates of the measured system and the pointer states of the apparatus.

\vn It may be mentioned that even this quantum correlation is only a notional one, since it implies an indirect correlation between pointer states in virtue of the system-apparatus entanglement, which is instantaneously erased by the process of environmental decoherence. This finally leads to a joint state of the measured system and the measuring apparatus where there remains only the classical correlation between the two. Each specific eigenstate is now {\it classically} correlated with a corresponding pointer state. The resulting state of the system-apparatus composite is one where the read-out process of the pointer state has not occurred, and corresponds to a probability distribution over the eigenvalues consistent with the Born rule. If now the pointer value of the apparatus is read out then some particular eigenvalue of the observable under measurement is obtained as the measurement result, and the relevant eigenstate is obtained as the reduced state of the system, thereby reproducing wave function collapse.

\vn In summary, {\it two} processes take place side by side,---the measurement interaction and the environment-induced decoherence of the state of the measuring apparatus. We recall that the apparatus is a classical object---a limiting instance of a quantum mechanical system---that can be looked upon as a composite system, with subsystems described by the collective macroscopic variables on the one hand, and the microscopic ones on the other. The two processes are of a contrary significance---one building up quantum correlations between the measured system and the measuring apparatus, and the other erasing the correlations as soon as those are generated, so that only the classical correlation remains.         

\subsection{The system-apparatus entanglement: a general principle}\label{general-sec}

\vn Assuming that the processes of measurement interaction and environmental decoherence occur independently and exclusively of each other, what is the measure of quantum correlations generated in the former and, equivalently, the extent of correlations erased in the latter? The answer to this can be expressed as a general principle governing quantum measurement processes: {\it as a result of the environmental decoherence the system-apparatus state goes over to the {\it nearest} separable state}, i.e., it is only the quantum correlations (the system-apparatus entanglement generated by means of the measurement interaction) that gets erased (recall the definition of entanglement measure mentioned in sec.~\ref{technical-sec}).
   
\vn\begin{quote}\begin{scriptsize}The initial system-apparatus state is one without any correlation (quantum or classical) and the pre-readout state is a separable one where, once again, there is no quantum correlation. Incidentally, in the transformation from the pre-readout to the post-readout state it is only the classical correlation that is erased in the readout process. \end{scriptsize}\end{quote}

\vn It turns out that, in virtue of the macroscopic nature of the measurement apparatus, the measure of quantum correlations erased in the process of environment-induced decoherence is {\it infinitesimally small} (\cite{Lahiri3}) compared to the classical correlations that remain in the pre-readout state. In the limit of an infinitely large measurement apparatus, the measure of quantum correlations erased in the environmental decoherence process goes to zero in comparison with the classical correlation that remains, while {\it the decoherence time also goes to zero} at the same time. This provides the basis of the measurement postulate where the wave function is assumed to collapse instantaneously as a measurement is performed.

\section{Quantum theory and reality}\label{reality-sec}

\nn Theories don't bear the responsibility of providing a `true' description of the workings of the world---of the noumenal world, that is. And, quantum theory is no exception. Like all other theories, it is a conceptual construct aimed at explaining observed regularities in the behavior of entities in our phenomenal world and that too only in a certain {\it context}. It is a remarkable theory precisely because it has been produced by the operation of the equally remarkable {\it inferential ability} of humankind, operating in the minds of its originators.

\vn The context in which quantum theory works as a true description of the behavior of entities in the phenomenal world is set, on the one hand, by the spatial and temporal scales of atomic and molecular processes, and by the {\it Planck scale} on the other. It is commonly agreed upon that quantum theory is more fundamental than the classical one. However, this is not to be interpreted as implying that the former can {\it replace} the latter. Theories are of contextual validity and form a complex mosaic in our conceptual space---there exist {\it common borders} between theories with overlapping domains of applicability. The asymmetric relation between successively emerging theories manifests itself in a neighborhood of their common border, where the latter is generally not a precisely defined one. Across the border, results of the classical theory can be explained and understood in terms of quantum principles, but the converse does not hold. Indeed the relation between successive theories is often complex---for instance the relation between quantum mechanics and classical mechanics or, say between wave optics and ray optics involves a limiting transition of the {\it singular}  type where quantitative predictions in the broader theory are related to those of the narrower one by means of {\it asymptotic series} (in this context, see~\cite{Berry}).

\vn The fact that the domain of validity of quantum theory (including quantum field theory) is limited by the Planck scale is consistent with the observation made above that the decoherence time for the measurement apparatus in a quantum measurement is possibly less than the Planck time where the decoherence by means of field fluctuations is taken into account. In other words, across the Planck scale, a new space-time domain is likely to be opened up where quantum theory is to be succeeded by a new and revised theoretical framework in terms of which the present day concepts of quantum theory will appear in a new light---on the other hand, the present day concepts will turn out to be inadequate in explaining or comprehending the emerging ones. For instance, the gyromagnetic ratio of the electron will possibly acquire a very small correction while the concept of the gyromagnetic ratio itself will perhaps appear in a new light---indeed, the electron itself will perhaps appear in a new incarnation in the emerging theory (in quantum field theory, the electron already acquires a radically altered interpretation, namely, as a quantum mechanical state of a field). It is, of course, an entirely different matter as to whether that new space-time domain will actually be opened up for us since the required technological, financial, and human resources will almost certainly prove to be too prohibitive for mankind, already precariously close to self-destruction.

\vn Assuming that the conditions for the emergence of a new domain of experience beyond the Planck scale are met with (a big assumption indeed) the present `incompleteness' of quantum theory will, in a manner of speaking, be rectified (and, of course, new areas of incompleteness will show up in a new context) but not in a manner that would prove satisfying to those stalwarts of the theory who wanted to have objectively existing `elements of reality', where observable quantities pertaining to an entity would have pre-existing values regardless of measurements being performed on those. The theory will, on the other hand, be made more complete by way of corresponding to an {\it expanded} domain of experience where new concepts will emerge as they have emerged in the past, making for a more enriched mosaic of theories with which we will be able to explain the way our newly emerged phenomenal world behaves.  

\vn The issues of `objectivity' and `truth' are vexed ones since their {\it context-dependence} is often overlooked. A theory cannot claim unconditional and absolute objectivity precisely because it applies to some particular domain of experience, thereby ignoring a great many correlations that are likely to influence the behavior of entities of interest beyond certain spatial and temporal scales. Across those relevant scales, {\it remote causes} assume significance, i.e., effects of correlations that do not show up within those scales, get revealed as new {\it modes} of behavior. Since a theory is designed to explain phenomena within some particular domain of experience, it necessarily ignores remote causes beyond that domain, and hence possesses objectivity only within the context to which it pertains. Absolute objectivity is a Platonic ideal that applies only to theories purporting to apply to the noumenal world which is a complex and integral whole involving all possible correlations between all possible putative entities that make up that world. In this context, `truth' is an even more problematic concept in that it is based on the idea of objectivity on the one hand and on that of {\it acceptability} or {\it justification} on the other. In a mathematical theory, justification is achieved by means of {\it proof} and it is commonly the case that all correct statements are provable. This, however, is not the case in the sciences where justification rests on {\it evidence}, and confirmation against evidence is never a closed process independent of human judgment.

\vn\begin{quote}\begin{scriptsize}A widely acclaimed discourse on truth is that by Alfred Tarski (see, for instance,~\cite{Wolenski}), and is referred to as the {\it semantic theory} of truth. It essentially defines the truth of a sentence in some language in terms of a correspondence between the structure of the sentence and a statement involving elements of a set (a `domain') and functions and relations defined in that set. This theory points towards one important requirement for a definition of truth to be meaningful, namely, the requirement that a true statement has to have {\it objective} validity, i.e., validity in some specified universe of discourse (as in a set with functions and relations defined in it). The question now arises as to how is the objective validity to be determined? In real life, this leads to the issue of {\it acceptability} where, as stated above, in mathematics, acceptability depends on the existence of a {\it proof}. On the other hand, in the sciences, and even more in {\it social} life, acceptability depends on confirmation by evidence. However, whether or not an evidence confirms a statement is itself a matter of theory (for instance, in social life, it depends to a large extent on preconceived notions). In other words, the {\it subjective} and the objective aspects of truth bear a complex relation to each other. \end{scriptsize}\end{quote}

\vn In view of all this, a scientific theory provides an objectively true description of the behavior of entities only within some particular context within our phenomenal world or, equivalently, within some particular {\it model}---a chunk scooped out from the phenomenal world, for which the infinitely complex collection of remote causes (i.e., correlations arising from entities and events that the model does not explicitly account for) are taken into consideration in an approximate manner by means of the context of the model (for instance, the initial and boundary conditions in the case of a model defined in terms of a set of partial differential equations). Within the limits of the model, more or less transparent and specific criteria for the acceptability of the theory can be formulated.

\vn\begin{quote}\begin{scriptsize}In the statistical description of complex systems, one often comes across probability distributions of the so-called `fat-tailed' type (see~\cite{Thurner}), represented by power-law probability density functions. These distributions typically arise as a consequence of the operation of the remote causes.\end{scriptsize}\end{quote} 
 
\vn In this sense, quantum theory lacks unconditional objectivity and truth as all other scientific theories do. And in this sense, it is in need of revision (into a more complete theory) beyond the context set by the Planck scale. The `remote causes' that quantum theory fails to account for are the pervasive field fluctuations across the Planck scale that result in an almost instantaneous environmental decoherence of states of a classical object---in particular, for the `measurement apparatus' used for a quantum mechanical system for preparing it in some (completely or partially known) state.

\vn\begin{quote}\begin{scriptsize}It is indeed intriguing that a quantum measurement does not measure or reveal a pre-existing value, but accomplishes the task of preparing a system in some particular state, depending on the value obtained in the process. \end{scriptsize}\end{quote} 

\subsection{The ontology of quantum theory}\label{ontology-sec}

\nn As in the case of all theories of the world, quantum theory carries its own ontology, where the respective ontologies of all our current scientific theories considered severally form a consistent mosaic, the latter as a whole constituting the {\it world view} that science provides us with. The consistency between parts of the mosaic is an intriguing thing since there remain apparent incompatibilities, underlying which there is a complex transition from one tile of the mosaic to its adjacent one.

\vn\begin{quote}\begin{scriptsize}The statement that a theory carries its own ontology is to be interpreted in the sense that the theory in question posits the sets of entities that it supposes to exist and the interactions among these entities that it proposes as being relevant in explaining various events within the context it applies to where the context, defined more or less precisely, specifies a model.\end{scriptsize}\end{quote}                    

\vn The fundamental ontology that distinguishes quantum theory is the following: entities (particles and fields) of the world are linked together in a pervasive network of correlations that can be described as one based on (phase dependent) complex amplitudes rather than on probabilities, where the correlations (quantum entanglement) are shared {\it globally} and are generated at all levels, even down to the one of field fluctuations at the shortest length and time scales.

\vn It is the globally shared aspect of the quantum correlations that characterizes {\it environmental decoherence}. More specifically, decoherence occurs in virtue of a global sharing of entanglement among environmental degrees of freedom, where the latter arise mostly in the form of oscillatory modes of fields pervading through space and time. Quantum theory without due recognition of this global sharing of phase correlations (i.e., the familiar version based only on the local correlations produced by means of direct interactions between systems) lacks coherence in the form of an incompatibility between the wave function collapse postulate and the rest of the quantum principles. Compatibility is restored (though only notionally, see below) as one recognizes that the global sharing of correlations is effectively random in nature and cause an extremely rapid removal of locally generated entanglement in the case of classical objects. It is the same process of erasure of locally generated phase correlations that explains as a limiting case the probability-based correlations among pre-existing values of `elements of reality' that characterize classical objects---ones that conform to the principle of `objectivity'. 

\vn However, the `limiting case' involves problems of a foundational nature since it calls upon us to look beyond the Planck scale. If and when the boundary relating to the Planck scale is crossed, quantum theory (and quantum field theory too) will be replaced with a broader theoretical framework that will bring in its own broader ontology. 

\vn\begin{quote}\begin{scriptsize}Donald Hoffman, in an insightful work (\cite{Hoffman}), asks the question `What could it mean to claim that no tomato is there when I don’t look?', and makes a case that the question isn't as silly as it looks at first sight. Hoffman's response to the question is indeed thought-provoking, though it is not quite in tune with the position the present paper is based on. For Hoffman seems to conflate here the noumenal and the phenomenal. The tomato is very much an entity in our phenomenal world, in which its very appearance depends on our perception. Beyond perception, what remains is noumenal `stuff' covering a large number of spatial and temporal scales. When we do look, a vast amount of information is filtered out from the noumenally existing reality (depending on how our mind works when confronted with that reality) so as to create the appearance of the tomato. If we look at a tomato thrown in the air, shut off all perception, and then look again, it will be found in exactly the same condition that a solution of its equations of motion would predict, implying that the phenomenal existence continues to be there all along. The phenomenal is nothing but a projection (a filtered-out appearance) generated from the noumenal, inextricably tied with the latter. Of course, all said and done, the response to the question `what happens to a tomato when we don't look at it?', depends on the meaning we attach to the question itself, which quite naturally would differ from one person to another. Regardless, Hoffman's take on quantum reality is very similar to ours, especially so in respect of the assertion that the response to `what happens to a quantum mechanical system when we don't make a measurement on it?' is fundamentally similar to the above question concerning the tomato.

\vn Incidentally, it is a keen insight of Hoffman's that tells us that our conception of space-time is very much an idea that pertains to the phenomenal reality---an insight that is in consonance with the Kantian tradition. Beyond the phenomenal, one is in the dark as to what happens to space-time or to the idea that the world evolves {\it in} space-time. However, it is certainly legitimate to say that our conception of space-time and of an evolution within the framework of space-time is a projection from the noumenal. Allowing ourselves to project backward just for the sake of it, we can notionally speak of the noumenal space-time and the noumenal evolution. Even so, we can, with less agnosticism, think of how the conception of space time (and of spatio-temporal evolution) is going to change as the current quantum theory is revised to something in keeping with future observations across the Planck scale. An entire batch of the present generation of theorists are engaged in speculations in this terrain.
\end{scriptsize}\end{quote}

\subsection{Theory revision: general considerations}\label{revision-sec} 

\nn\begin{quote}\begin{scriptsize} Items 7,8,9 of sec.~\ref{mind-sec} provide a general background to the complex issue of theory revision, to which we add here a few incidental comments in the context of the present essay. This will entail some repetitions too from earlier parts of it. We once again refer to~\cite{Lahiri1} for background and to further details. \end{scriptsize}\end{quote}

\vn Theories are conceptual constructs and are essentially in the nature of bundles of beliefs by means of which we explain observed events in the phenomenal world of our experience and---what is of vital relevance---act back on that world, based on anticipations of future events. It is the inferential ability of the human mind (see~\cite{Lahiri1}) that makes our theories latch on to reality---fragmentary patches of the noumenal reality caught from various different perspectives. In order to be effective in helping us in our journey in life, the theories have to connect with reality, but that does not make them approximations to a true description of the noumenal reality---recall that the latter can only be described as an integral whole since any cross-section of it can only pertain to the phenomenal reality that we assemble with the help of our interpretations.

\vn This apparently runs counter to what is referred to as {\it scientific realism}. However, the position we adopt in this paper is that scientific realism can only mean the irreducible meta-induction that the noumenal world exists independently of our senses and is the ultimate source of all our perceptions, interpretations, and theories---it constitutes the metaphysics this essay rests upon. Accordingly, the view that theories are conceptual constructs pertaining to models in our phenomenal world does not run counter to this central tenet of scientific realism.

\vn One can go further and state that theories emerging through successive revisions do not constitute `approximations' to some grand principle inherent in Nature, but are nothing more nor less than revisions in our belief system. They occur piecemeal, as and when our domains of experience undergo expansion, being dictated by circumstances in life---mostly fortuitous and unpredictable. Successively emerging theories bear a relation of incommensurability with one another---concepts in the earlier theory are comprehensible in terms of the one emerging later, but many of those in the latter are not explicable in terms of the former. 

\vn\begin{quote}\begin{scriptsize}The incommensurability arises because of the {\it multi-layered} structure (see~\cite{Thurner}) of the complex network of our concepts, where the latter are connected by links representing relations of varied kinds (such as the relations of association and implication). The network undergoes a radical restructuring with the emergence of a new theory, whereby new layers of links are established while some earlier links get removed and some others get modified. It is the newly established links that renders concepts in the emerging theory incomprehensible in terms of those in the earlier one (see~\cite{Lahiri1}) while the earlier concepts are mostly explicable in terms of the later ones.\end{scriptsize}\end{quote}

\subsection{What will quantum theory be revised to?}\label{revised-sec}

\nn Can we foresee as to how and when the present day quantum theory will be revised? Obviously, No! But the general features of theory revision can safely be assumed to apply. The newly emerging theory, {\it if and when it comes}, will be a conceptual construct pertaining to an expanded terrain of experience, where it will explain a number of newly observed phenomena and will successfully predict a yet larger body of phenomena to be subsequently unearthed and, {\it additionally}, will throw the present day concepts of quantum theory in a new light. The novel concepts of the emerging theory, however, will be beyond the reach of these present day concepts. In this sense, relation between the present day theory and the emerging one will be {\it asymmetric} and {\it incommensurable}. The new theory will arise in consequence of a set of creative inferences of the human mind, and will correctly latch on to features of an expanded and enlarged phenomenal reality---the reality that we perceive by means of innumerable signals received from the noumenal world. While the phenomenal is a projection from the noumenal, the reverse projection is not in the realm of feasibility.

\vn\begin{quote}\begin{scriptsize}A child is suddenly flooded by a dread that her mother is not present in her nearabouts and starts crying. She arrives at the inference without conclusive evidence and yet her inference does, in all probability, apply to the reality around her. Human inference, which is mostly inductive in nature, likewise results from an ability of the human mind to reconstruct reality, though often in a skewed fashion.\end{scriptsize}\end{quote}    

\vn Our metaphysical position---of a broadly Kantian nature---implicitly tells us that the emerging theory will be a far cry from a {\it foundational} theory of reality, or a {\it theory of everything}. There can be no such foundational theory because reality---the noumenal reality, that is,---is an infinite dimensional complex universe in a constant state of complex evolution and is responsible for the phenomenal reality that we perceive. The latter is constructed piecemeal by means of a sporadically expanding terrain of experience, and a concomitant expansion---equally irregular in its progress---of the complex mosaic of our theoretical constructs. Theories often get revised radically. The successive revision of theories is certainly `progressive' in the sense that they apply to a constantly expanding universe of our phenomenal experience but these do not represent a progressively increasing `truth content' of our description of nature because the latter is an infinitely complex whole, and because the concepts that emerge successively bear no simple and monotonic relation to one another and do not conform to an overall pattern.

\vn Incidentally, it is often asserted in connection with the ontology of quantum mechanics that `reality is non-local' or `reality is contextual', referring thereby to the non-locality or the contextuality (see, for instance,~\cite{Mermin}) that emerges as a necessary consequence of Bell's theorem or the Kochen-Specker theorem, as the case may be. Such an interpretation, however, is not warranted since the non-locality or the contextuality in question is found to be a necessary feature of a (hidden-variable) theory that tries to explain measurement results {\it in classical terms}. Quantum theory in itself does not attach any such tag of non-locality or contextuality to `reality' since the one single ontological posit of quantum theory is that of globally shared phase-dependent correlations (i.e., entanglement) among states of entities. It is precisely the {\it global} aspect of the correlations, along with the universal principle characterising quantum measurements (section~\ref{general-sec}) that bring in the features of contextuality and non-locality when a classical explanation of quantum mechanical measurement results is attempted.

\vn Indeed, contextuality in a broad sense is a general characteristic of any and every theory pertaining to reality, and is not of specific relevance to quantum theory alone. This is because of the fact that a theory pertains to some particular model or a part of reality and applies in the context of that model alone where the effect of correlations arising from the remaining part of reality is either ignored or accounted for only partially.

\pagestyle{plain}

\end{document}